\documentclass[aps,prl,twocolumn,groupedaddress,preprintnumbers,floatfix,10pt,nofootinbib]{revtex4-1}
\pdfoutput=1
\usepackage[bookmarks=false]{hyperref}
\usepackage{amsmath}
\usepackage{graphicx}
\usepackage{xcolor}

\newcommand{\refcite}[1]{Ref.~\cite{#1}}
\newcommand{\refscite}[1]{Refs.~\cite{#1}}
\newcommand{\eq}[1]{Eq.~\eqref{eq:#1}}
\newcommand{\eqs}[2]{Eqs.~\eqref{eq:#1} and \eqref{eq:#2}}

\newcommand{\fig}[1]{Fig.~\ref{fig:#1}}

\newcommand{\tab}[1]{Table~\ref{tab:#1}}

\newcommand{\nn}{\nonumber}

\newcommand{\abs}[1]{\lvert#1\rvert}

\newcommand{\ord}[1]{\mathcal{O}(#1)}
\newcommand{\Ord}[1]{\mathcal{O}\bigl(#1\bigr)}

\newcommand{\df}{\mathrm{d}}

\newcommand{\eps}{\epsilon}
\newcommand{\la}{\lambda}
\newcommand{\w}{\omega}

\newcommand{\cF}{\mathcal{F}}
\newcommand{\cH}{\mathcal{H}}
\newcommand{\cI}{\mathcal{I}}

\newcommand{\cO}{\mathcal{O}}

\newcommand{\alphas}{\alpha_s}

\newcommand{\lqcd}{\Lambda_\mathrm{QCD}}
\newcommand{\MSbar}{$\overline{\text{MS}}$}

\newcommand{\kt}{{\vec k}_T}
\newcommand{\pt}{{\vec p}_T}
\newcommand{\qt}{{\vec q}_T}
\newcommand{\PTb}{\vec P_{hT}}
\newcommand{\phiH}{\phi_h}       
\newcommand{\phiS}{\phi_S}       

\newcommand{\GeV}{\,\mathrm{GeV}}

\newcommand{\fb}{\,\mathrm{fb}}

\newcommand{\acop}{\mathrm{acop}}

\newcommand{\EIC}{{\mathrm{EIC}}}

\newcommand{\rest}{\mathrm{rest}}
\newcommand{\NP}{\mathrm{NP}}

\allowdisplaybreaks[2]

\providecommand{\sectionPaper}[1]{\section{#1}}
\providecommand{\headingAcknowledgments}{\paragraph{Acknowledgments}}
\providecommand{\app}[1]{Supplement~\ref{app:#1}}

\begin{document}


\preprint{\vbox{\hbox{MIT-CTP 5462}}}

\title{A Better Angle on Hadron Transverse Momentum Distributions at the EIC}

\author{Anjie Gao}%
\email{anjiegao@mit.edu}%

\author{Johannes K.\,L.\ Michel}%
\email{jklmich@mit.edu}%

\author{Iain W.\ Stewart}%
\email{iains@mit.edu}%

\author{Zhiquan Sun}%
\email{zqsun@mit.edu}%
\affiliation{Center for Theoretical Physics, Massachusetts Institute of Technology, Cambridge, MA 02139, USA}%

\date{April 14, 2023}

\begin{abstract}
We propose an observable
$q_*$ sensitive to transverse momentum dependence (TMD) in
$e N \to e h X$, with
$q_*/E_N$ defined purely by lab-frame angles.
In 3D measurements of confinement and hadronization this
resolves the crippling issue of accurately
reconstructing small transverse momentum
$P_{hT}$.
We prove factorization for
$\mathrm{d} \sigma_h / \mathrm{d}q_*$
for
$q_*\ll Q$ with standard TMD functions,
enabling
$q_*$ to substitute for
$P_{hT}$.
A double-angle reconstruction method is given which is exact to all orders in QCD for
$q_*\ll Q$.
$q_*$
enables an order-of-magnitude improvement in the expected experimental resolution at the EIC.

\end{abstract}

\maketitle

\sectionPaper{Introduction}
\label{sec:intro}

A deeper understanding of the emergent properties of the nucleon, such as
confinement and hadronization,
has been a frontier of nuclear and particle physics research
since the inception of Quantum Chromodynamics (QCD) five decades ago.
An important one-dimensional view of the nucleon is provided by the  deep-inelastic scattering (DIS) process
$e^-(\ell) + N(P) \to e^-(\ell') + X$,
where the scattering is mediated by an off-shell photon
of momentum $q = \ell - \ell'$ (with $Q^2 \equiv -q^2 > 0$).
Confinement is probed by measurements of $x = Q^2/(2 P \cdot q)$,
the momentum fraction carried by the colliding parton inside the nucleon $N$.
A more intricate view is obtained by identifying a hadron $h$ in semi-inclusive DIS (SIDIS),
$e^-(\ell) + N(P) \to e^-(\ell') + h(P_h) + X$.
Here measurements of the longitudinal momentum fraction $z = (P \cdot P_h)/(P \cdot q)$ that the hadron retains when forming from the struck quark give insight into the complex dynamics of hadronization.
Measuring the hadron's \emph{transverse} momentum $\vec{P}_{hT}$ relative to $\vec{q}$
gives access to a three-dimensional view of the confinement and hadronization processes for $N$ and $h$,
together with spin correlations that probe these processes.

The region most sensitive to these dynamics occurs for small transverse momentum, $P_{hT} \sim \lqcd \ll Q$, where $\lqcd$ is the QCD confinement scale.
Here the cross section obeys a rigorous factorization theorem~\cite{Collins:1350496}, with the confinement and hadronization dynamics encoded in universal transverse momentum-dependent (TMD) parton distribution functions (PDFs) and fragmentation functions (FFs).
SIDIS cross sections have been extensively studied experimentally at HERMES\cite{Airapetian:2012ki,HERMES:2019zll,HERMES:2020ifk},
COMPASS\cite{Alekseev:2008aa,Aghasyan:2017ctw,Parsamyan:2018ovx},
RHIC\cite{Aschenauer:2015eha,Adamczyk:2015gyk}, and
JLab\cite{CLAS:2003qum,CLAS:2017yrm,CLAS:2021jhm}
and together with the Drell-Yan process
have enabled extractions of TMD PDFs and FFs (TMDs) by various groups, e.g.\ \cite{Scimemi:2019cmh,Bacchetta:2019sam,Bury:2021sue}.
A key scientific goal of the upcoming Electron-Ion-Collider (EIC)\cite{AbdulKhalek:2021gbh} is to study SIDIS with enormous beam luminosities to determine TMD PDFs and FFs with unprecedented precision. Progress has also been made towards  calculations of TMD PDFs from lattice QCD~\cite{Ebert:2018gzl,Shanahan:2020zxr,Shanahan:2021tst,Schlemmer:2021aij,Li:2021wvl,LPC:2022ibr}.

A key challenge in experimental studies of TMDs is
that measurements of $\vec{P}_{hT}$
require reconstructing the photon momentum (or Breit frame)
to great accuracy to avoid loss of precision on $P_{hT} = \abs{\vec{P}_{hT}} \ll Q$.
A misreconstruction of $\vec{\ell}'$ by $\ord{\Delta}$ leads to a misreconstruction of $\vec{q}$ and therefore $\vec{P}_{hT}$ by $\ord{\Delta}$, which is a large uncertainty for $P_{hT}\ll Q$.
For example, for a nominal measurement at $P_{hT}/z = 1\,{\rm GeV}$ with $Q=20\,{\rm GeV}$, a typical detector resolution of $\Delta=0.5\,{\rm GeV}$ leads to a 50\% uncertainty.
This puts in peril the EIC physics program to unveil the dynamics of hadronization and confinement in the kinematic region with the largest sensitivity.

In this paper we construct a novel SIDIS observable, $q_*$,
designed to be maximally resilient against resolution effects while delivering the same sensitivity to TMD dynamics as $\vec{P}_{hT}$.
The key insight is that while the magnitude of the electron and hadron three momentum
is subject to limited detector resolution,
modern tracking detectors deliver near-perfect resolution on the \emph{angles} of charged particle tracks.
We will therefore construct $q_*$ to satisfy the following three criteria:
(i)~it is purely defined in terms of lab-frame angles and the beam energies;
(ii)~at small values $q_* \ll Q$, the differential cross section $\df \sigma / \df q_*$,
including spin correlations, still satisfies a rigorous factorization theorem
in terms of the standard TMD PDFs and FFs;
(iii) it does not dilute the statistical power of the available event sample.
Our construction is inspired by, but features key differences to,
the Drell-Yan $\phi^*_\eta$ observable in hadron-hadron collisions~\cite{Banfi:2010cf},
which has enabled tests of perturbative QCD from permil-level $Z$-pole measurements
at the Tevatron and LHC~\cite{D0:2010qhp,LHCb:2012gii,ATLAS:2015iiu,LHCb:2015okr,ATLAS:2019zci,CMS:2022ubq}.

Below we define $q_*$ in detail,
prove the factorization theorem for $q_*$ with standard TMDs,
and evaluate the expected detector resolution,
statistical power, and resilience against systematic biases of $q^*$ versus $\vec{P}_{hT}$.

\sectionPaper{Constructing $q_*$}
\label{sec:constructing_qstar}

\begin{figure}[t!]
\includegraphics[width=\columnwidth]{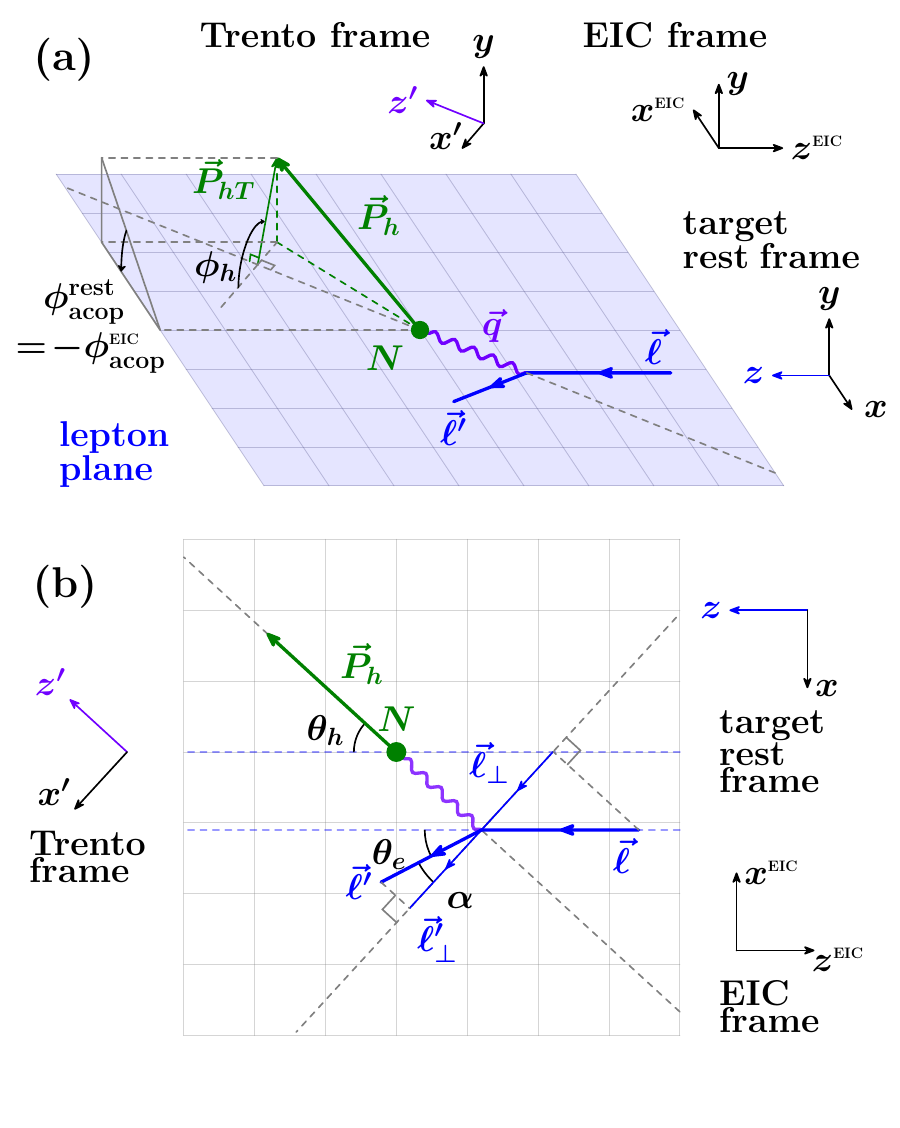}
\vspace{-0.5in}
\caption{%
(a) Definition of $\phi_\acop^\rest$ needed to construct $q^*$.
Momenta are not to scale.
We define the conventional Trento frame \cite{Bacchetta:2004jz} for SIDIS,
as well as the target rest frame and the EIC frame.
$\phi_{h}$ is the azimuthal separation between $\vec P_{h}$ and $\vec \ell$ in the Trento frame.
The acoplanarity angle in the target rest frame is $\phi_\acop^\rest \equiv \pi - \Delta\phi^\rest$,
where $\Delta\phi^\rest$ is the azimuthal separation between $\vec P_{h}$ and $\vec\ell^{\,\prime}$.
(b) In-plane geometry for leading-power kinematics $P_{h,\perp}/Q \ll 1$,
where we can approximate that $\vec P_{h}$ is along the same direction as $\vec q$.
This geometry yields the double-angle formulas in \eq{dbleangle}
$Q$, $x$, $y$ with angular measurements.
}
\vspace{-0.2in}
\label{fig:kinematics}
\end{figure}

Consider the target rest frame shown in \fig{kinematics}a where the nucleus $N$ is at rest
and the $z$-axis is along the incoming lepton beam.
The lepton momenta $\vec \ell$ and $\vec\ell'$ define the lepton plane as the $x$-$z$ plane.
We wish to take advantage of the high-precision reconstruction
of polar angles (rapidities) and azimuthal angles in the EIC lab frame.
Here we give results in terms of
EIC frame rapidities in the light target mass limit $M \ll Q$,
with full $M$ dependence in \app{finite_mass}.
The acoplanarity angle
in the target rest frame, $\phi_\acop^\rest$, is defined by  $\tan \phi_\acop^\rest = - P_{h,y}/P_{h,x}$,
where $P_{h,x}$ and $P_{h,y}$ are components of $P_h$.
From \fig{kinematics}a,
it is obvious that $\tan \phi_\acop^\rest$ $\propto \sin \phi_h \, P_{hT}$,
where $\phiH$ is the azimuthal angle of $\vec P_h$ in the Trento frame.
We may thus use $\phi_\acop$
as a precision probe of the hadron transverse momentum $P_{hT}$.%
\footnote{Lab-frame acoplanarity angles are also useful as a measure of transverse momentum in jet production in DIS~\cite{Liu:2018trl, Liu:2020dct}.
Unlike this work, the observable of \refscite{Liu:2018trl, Liu:2020dct}
does not feature the same experimental improvements
since traditional jet axis reconstruction is not angular,
and it also has nonglobal logarithms
that are nonperturbative in the TMD region of interest.}

To work out the full relation between $\phi_\acop^\rest$ and $P_{hT}$, consider now the leading-power (LP) kinematics illustrated in \fig{kinematics}\,b,
where $\lambda \sim P_{hT}/(zQ) \ll 1$.
We find
\begin{align}\label{eq:tan_phi_acop}
\tan\phi_\acop^\rest
&= \frac{\sin \phi_h  \, P_{hT}}{zQ \sqrt{1-y}} + \ord{\la^2}
\,. \end{align}
We now wish to express $Q$ and $y = (P \cdot q )/ (P \cdot \ell)$ in terms of final-state angles
in the EIC frame,
which is defined by a $180^\circ$ rotation about our rest frame $y$ axis and then a boost along the $z$-axis,
so $\phi_{\acop}^\EIC = -\phi_{\acop}^\rest$.
From \fig{kinematics}b,
momentum conservation gives $\ell_{x'} = \ell'_{x'}$, $q_{x} = -\ell'_{x}$,
and $P_{hT} \ll Q$ implies $\theta_{h} + \theta_{e} + \alpha = \pi/2$.
We find $y = 1- \sin\theta_{h}/\cos\alpha$ and
$Q^{2} = (\ell^0_\rest)^{2} [\frac{\sin^{2}\theta_{e}}{\cos^{2}\alpha} - (1-\frac{\sin\theta_{h}}{\cos\alpha})^{2}]$.
Boosting
to the EIC frame:
\begin{align} \label{eq:dbleangle}
Q^2 \! &=\! (2P^0_\EIC)^{2} \frac{e^{\eta_e +\eta_h}}{1 + e^{\Delta \eta}} \!+\!\ord{\la}
\,,  \,\,\
y \!=\! \frac{1}{1+e^{\Delta \eta}} \!+ \!\ord{\la^2}
\,,
\nn\\
x &= \bigl[(2 P^0_\EIC)^{2}/s \bigr] \, e^{\eta_e+\eta_h}+{\cal O}(\lambda)\,,
\end{align}
where $\eta_i$ are the EIC frame pseudorapidities of the outgoing lepton $i=e$ and hadron $i=h$,
$\Delta \eta \equiv \eta_h - \eta_e$, and $s = (P + \ell)^2$.
This construction agrees with the double-angle formula in \refcite{Bentvelsen:1992fu}.
However, Ref.~\cite{Bentvelsen:1992fu}
uses the struck quark angle in a tree-level picture, while
our \eq{dbleangle} uses the hadron angle
and holds to all orders in $\alpha_{s}$, and up to power corrections in $P_{hT}/(zQ)$, which controls the distance to the Born limit.
The ${\cal O}(\lambda)$ corrections to \eq{dbleangle} are given in \app{pc_dbleangle}.

To exploit the proportionality in \eq{tan_phi_acop} to probe $P_{hT}$,
we define an optimized observable:
\begin{align}\label{eq:qstar}
q_* &\equiv 2 P^0_\EIC \frac{e^{\eta_h}}{1+e^{\Delta \eta}}
\tan \phi_\acop^\EIC
\,.\end{align}
Expanding in $P_{hT}\ll zQ$ it has a simple LP limit
\begin{align}\label{eq:qstar_lp}
q_* \stackrel{\text{LP}}{=} - \sin \phi_h \frac{P_{hT}}{z} \,.
\end{align}
Thus for TMD analyses, $Q^2$, $x$, $y$, and $q^*$ can all be measured from the beam energy $P^0_\EIC$ and angular variables.
We may also define a dimensionless variable,
\begin{align}
\phi^*_\mathrm{SIDIS}
= \sqrt{\frac{e^{\Delta \eta}}{1+e^{\Delta \eta}}}
\tan \phi_\acop^\EIC
= \, \frac{q_*}{Q}
+ {\cal O}(\lambda^2)
\,.\end{align}
This is analogous to the setup for the $\phi^{*}_{\eta}$ observable in Drell-Yan~\cite{Banfi:2010cf}.
We expect the purely angular observables $q_{*}$ and $\phi^*_\mathrm{SIDIS}$
to be measured to much higher relative precision compared to the transverse momentum $P_{hT}$.

\sectionPaper{Factorization}
\label{sec:factorization}

Consider the standard factorization theorem for polarized SIDIS~\cite{Ji:2004wu,Ji:2004xq,Bacchetta:2006tn,Collins:1350496},
\begin{align} \label{eq:leading_SIDIS_pol}
& \frac{\df\sigma}{\df x \, \df y \, \df z \, \df^2\PTb}
=  \sigma_{0}
\Bigl\{
W_{UU,T} + \lambda_e S_L \sqrt{1-\eps^2} \, W_{LL}
\nn \\ & \:\;
+ \eps \cos(2\phiH) W_{UU}^{\cos(2\phiH)}
+ S_L \eps \sin(2\phiH) W_{{UL}}^{\sin (2\phiH)}
\nn \\& \:\;
+ S_T \sin(\phiH \! -\! \phiS) W_{UT,T}^{\sin(\phiH \!- \phiS)}
+ \eps S_T \Bigl[ \sin(\phiH \! +\! \phiS)
\nn \\& \:\; \:\;
\times W_{UT}^{\sin(\phiH \!+ \phiS)} \!\!
+ \sin(3 \phiH \! - \! \phiS) W_{UT}^{\sin(3 \phiH \! -\phiS)} \Bigr]
\nn \\& \:\;
+ \lambda_e S_T \sqrt{1-\eps^2} \cos(\phiH \! - \! \phiS) W_{LT}^{\cos(\phiH \! -\phiS)}
\Bigr\}
\,,\end{align}
where $\sigma_{0} \equiv \alpha^{2} \pi y \kappa_{\gamma}/[ z Q^{2} (1-\eps)]$,
$\alpha$ is the fine-structure constant,
$\lambda_e$ is the lepton beam helicity,
$S^\mu = (0, S_T \cos \phi_S, S_T \sin \phi_S, -S_L)$
is the nucleon spin vector in the Trento frame~\cite{Bacchetta:2006tn},
$\eps = (1-y)/(1-y+y^2/2)$, and $\kappa_\gamma = 1$ for $M \ll Q$.
We have only kept the structure functions $W$
that are nonzero at leading power in $\la$.
They can be written in terms of the hard function $\cH (Q^{2})$
and various $b_{T}$-space TMD PDFs $\tilde g (x,b_T)$ and TMD FFs $\tilde D(z,b_{T})$~\cite{Boer:2011xd}:
\begin{align}\label{eq:fourier_conv}
W_{PP'}^{\rm ang} & \propto \cF \bigl[ \cH \, \tilde g^{(n)} \! \tilde D^{(m)} \bigr]
\\
&\equiv  2 z \int_0^\infty \! \frac{\df b_T b_T}{2\pi} \, \cI\bigl[\cH \, \tilde g^{(n)} \tilde D^{(m)} \bigr]
J_{n+m}(b_T P_{hT}/z)
\nn
\end{align}
with
$\cI [\cH \, \tilde g^{(n)} \tilde D^{(m)}]
\equiv (Mb_{T})^{n} (- M_{h} b_{T})^{m} \sum_{f} \! \cH_{f}
\tilde g_{f}^{(n)} \tilde D_{f}^{(m)}$,
where $f$ sums over quarks and antiquarks.
For example, $W_{UU}^{\cos(2\phiH)} =
-\cF \bigl[\cH \,\tilde h_1^{\perp(1)} \tilde H_1^{\perp(1)}\bigr]$,
where $\tilde h_1^{\perp(1)}$ and $\tilde H_1^{\perp(1)}$ are the Boer-Mulders~\cite{Boer:1997nt} and Collins~\cite{Collins:1992kk} functions.
For details on our  notation and conventions, see \refcite{Ebert:2021jhy}.

To compute the spectrum differential in $x, y, z$ and $q_*$,
we insert the leading-power measurement $\delta(q_* +\sin\phiH  P_{hT} / z)$ and analytically perform the integral over $\df^2 \vec{P}_{hT} = \df P_{hT} \, P_{hT} \df \phi_h$.
As an explicit example, we work out the contribution from $W_{UU}^{\cos(2\phiH)}$. Using \eq{fourier_conv}:
\begin{align} \label{eq:FactCos2phih}
&\int_0^\infty \!\!\! \df P_{hT} P_{hT} \! \!
\int_0^{2\pi} \!\!\!\!\! \df \phiH \,
\delta \Bigl( q_* \! + \! \sin \phi_h \frac{P_{hT}}{z} \Bigr)
\cos(2\phiH) W_{UU}^{\cos(2\phiH)}
\nn \\
&= -\frac{2z^{3}}{\pi} \int \! \df b_T \, \cI\bigl[\cH \,
\tilde h_{1}^{\perp(1)} \! \tilde H_{1}^{\perp (1)} \bigr]
\nn \\
&~\times \int_{0}^{2\pi} \!\!\! \frac{\df \phi_{H}}{\sin^2 \! \phiH} \,
\Theta\Bigl(-\frac{q_*}{\sin\phiH}\Bigr)
\cos(2\phiH) \, \frac{b_T \abs{q_*}}{2} \,
J_{2}\Bigl (\frac{b_{T} q_* }{\sin\phiH} \Bigr)
\nn \\
&= - \frac{2 z^{3} }{\pi} \int \df b_{T} \, \cI\bigl[\cH \,
\tilde h_{1}^{\perp(1)} \! \tilde H_{1}^{\perp (1)} \bigr]
\,\cos(q_* b_T)
\,.\end{align}
The $\phi_h$ integral,
which is specific to the structure function,
can only depend on $q_* b_T$ by dimensional analysis,
and in this case yields a simple $\cos(q_* b_T)$.

In total,
the LP cross section differential in $q_*$ is:
\begin{align} \label{eq:qstarfact}
&\frac{\df \sigma}{\df x \, \df y \, \df z \, \df q_*}
= \frac{ 2 z^{3}}{\pi} \sigma_0
\int_{0}^{\infty} \!\!\! \df b_{T} \,\Bigl\{
\cos(q_* b_T) \Bigl (
\cI \bigl[\cH \, \tilde f_{1} \, \tilde D_{1}\bigr]
\nn \\
& \quad
- \eps \, \cI\bigl[\cH \, \tilde h_{1}^{\perp(1)} \! \tilde H_{1}^{\perp(1)}\bigr]
+ \lambda_e S_{L}\sqrt{1\!-\!\eps^{2}} \, \cI\bigl[\cH \, \tilde g_{1L} \, \tilde D_{1}\bigr]
\Bigr)
\nn \\
& \:\;
+ \cos\phiS \sin(q_* b_T) \, S_{T} \Bigl(
\cI\bigl[ \cH \, \tilde f_{1T}^{\perp (1)} \tilde D_{1 }\bigr]
+ \eps \, \cI[\cH \, \tilde h_{1} \, \tilde H_{1}^{\perp (1)}]
\nn \\
& \quad
+ \frac{\eps}{4} \,  \cI\bigl[\cH \,
\tilde h_{1T}^{\perp (2)} \tilde H_{1}^{\perp (1)} \bigr]
\Bigr)
\nn \\
& \:\;
- \sin\phiS \sin(q_* b_T) \, \lambda_e S_{T} \sqrt{1-\eps^{2}} \, \cI \bigl[\cH \,  \tilde g_{1T}^{\perp(1)} \tilde D_{1}\bigr]
\Bigr\}
\,.\end{align}
We stress that the TMD PDFs and FFs $\tilde f_1, \tilde D_1, \tilde h_1^{\perp(1)}, \dots$
are the \emph{same} as in the standard factorization for the $P_{hT}$ spectrum and TMD spin correlations.
This is analogous to the role of the unpolarized~\cite{Banfi:2009dy, Banfi:2011dx} and Boer-Mulders~\cite{Ebert:2020dfc} TMD PDFs in the Drell-Yan $\phi^*_\eta$.
The factorization theorem can equivalently be written
in terms of momentum-space TMDs, see \app{momentum_space_factorization}.
Crucially, definite subsets of these TMDs contribute to the even and odd parts
of the spectrum under $q_* \to -q_*$.
The odd parts are accessible through
the asymmetry $\df \sigma(q_* > 0) - \df \sigma(q_* < 0)$.
Contributions can be further disentangled experimentally through their unique dependence
on $\lambda_e$, $S^\mu$ and $\eps$,
i.e., by taking asymmetries with opposite beam polarizations and by measuring cross sections as a function of $y$
.%
\footnote{We recommend reconstructing $S^\mu$
using a rotation by $\theta_h$ to maintain a purely angular measurement,
see \app{angular_target_spin}, which is justified at LP.
Note that the transversity and pretzelosity PDFs have a degenerate contribution $\eps S_T (h_1 + h_{1T}^\perp/4)$ to $q_*$ in \eq{qstarfact},
while the worm-gear $L$ function $h_{1L}^\perp$ drops out,
due to $q_*$ being even under $\phi_h \leftrightarrow \pi - \phi_h$.
Encouragingly, the subleading-power Cahn effect $\propto \cos(\phi_h)$,
which pollutes standard $\vec{P}_{hT}$, also drops out for the same reason.}
E.g., the double asymmetry for $q_* \to -q_*$ and $\lambda_e \to -\lambda_e$ as a function of $x$ and $\abs{q_*}$
gives direct access to the worm-gear $T$ function $\tilde g_{1T}^\perp(x, b_T)$.

\sectionPaper{Experimental sensitivity}
\label{sec:sensitivity}

\begin{figure}[t!]
\includegraphics[width=\columnwidth]{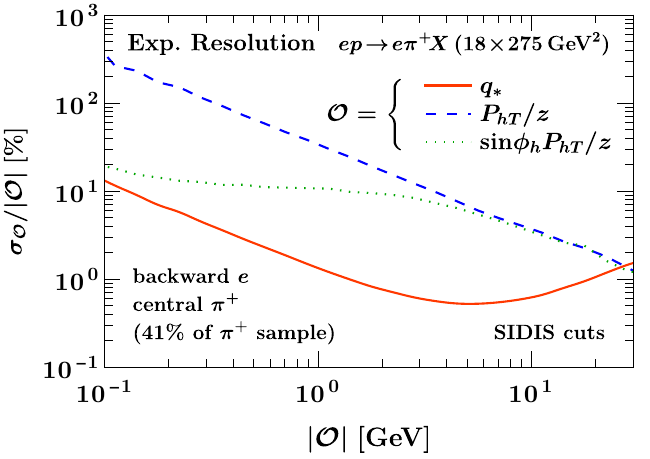}
\vspace{-0.2in}
\caption{%
Expected event-level detector resolution $\sigma_\cO$ for different SIDIS TMD observables $\cO$.
We show relative resolutions as a function of the magnitude of $\cO = q_*$ (solid red), $P_{hT}/z$ (dashed blue), and $\sin \phi_h P_{hT}/z $ (dotted green).
}
\vspace{-0.2in}
\label{fig:resolution_curves_bc}
\end{figure}

To show the improvement that $q_*$ makes for TMD analyses, we use \texttt{Pythia~8.306}~\cite{Bierlich:2022pfr} to simulate $e^- p \to e^- X$ at $18 \times 275 \GeV^2$,
disabling QED corrections.
We select on $h=\pi^+$ and apply the following cuts (``SIDIS cuts''):
\begin{align}
x &> 0.001
\,, \quad
&0.01 &< y < 0.95
\,, \quad
z > 0.05
\,, \nn \\
Q^2 &> 16 \GeV^2
\,, \;
&W^2 &= (P + q)^2 > 100 \GeV^2
\,.\end{align}
Scaled to an integrated EIC luminosity of $10 \fb^{-1}$,
this results in a sample with $N_{\pi^+} = 4.18 \times 10^8$.
We first assess the expected detector resolution
of $q_*$ compared to $P_{hT}$.
We apply Gaussian smearing
to the final-state electron and hadron momenta, assuming a tracking detector that matches
the
performance given in \refcite{AbdulKhalek:2021gbh}:
a resolution of
$\sigma_p/p = 0.05\% \, p/\!\GeV \oplus 0.5\%$
on the momentum $p = \abs{\vec{p\mkern-2mu}\mkern2mu}$ of charged particles in the central barrel region $\abs{\eta} < 1$,
$0.05\% \,p/\!\GeV \oplus 1\%$ in the inner endcap $1 < \abs{\eta} < 2.5$, and
$0.1\% \,p/\!\GeV \oplus 2\%$ in the outer endcap $2.5 < \abs{\eta} < 3.5$.
A particle is forward (backward) if it has $1<|\eta|<3.5$ and
$\eta > 0$ ($\eta < 0$).
We assume a fixed angular resolution of $\sigma_\theta = \sigma_\phi = 0.001$.
We ignore the electromagnetic calorimeter as its $e^-$ energy resolution
is expected to be a factor of two worse than the tracker~\cite{AbdulKhalek:2021gbh}.
Our key results for the
detector resolution on $q_*$
compared to $P_{hT}$ are shown in \fig{resolution_curves_bc}
for
the case of a backward $e^-$ and a central $h$,
which accounts for the largest share ($41\%$) of the event sample.
We see that $q_*$ improves over the resolution of $P_{hT}/z$
by {\it an order of magnitude} across the strongly confined TMD region $q_*, P_{hT}/z \lesssim 2 \GeV$.
Similar results are obtained for other detector regions, see \app{resolution_curves}.
It is interesting to compare $q_*$ to a direct measurement of
$\sin \phi_h P_{hT}/z$, to which it reduces at leading power.
The latter has improved resolution over $P_{hT}/z$
at small values thanks to picking up on the same acoplanarity of the event,
which is stable against the electron momentum resolution, but
cannot outperform $q_*$ since it is not
a pure angular variable.

\begin{figure}[t!]
\includegraphics[width=\columnwidth]{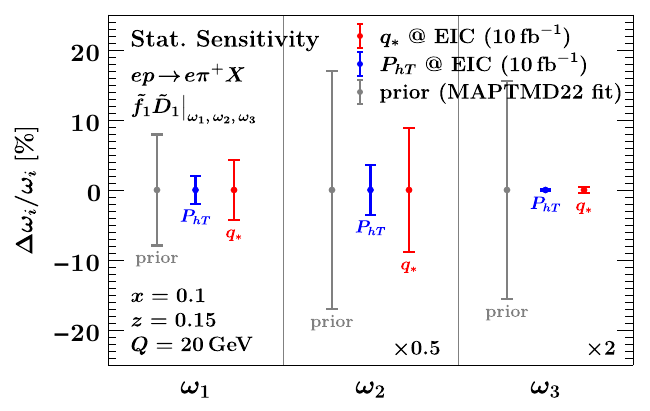}
\vspace{-0.2in}
\caption{%
Statistical sensitivity to TMD nonperturbative model coefficients
at the $10 \fb^{-1}$ EIC when measuring $P_{hT}$ (blue) or $q_*$ (red)
compared to the prior (MAPTMD22 fit~\cite{Bacchetta:2022awv}, gray).
Despite its superior resolution, $q_*$ enjoys comparable statistical sensitivity to $P_{hT}$.}
\vspace{-0.2in}
\label{fig:stat_sensitivity}
\end{figure}

To verify the statistical sensitivity of $q_*$ to TMD physics we perform a Bayesian reweighting analysis
of the unpolarized cross section $\df \sigma/(\df x \, \df z \, \df Q^2 \, \df \cO)$ for $\cO = P_{hT}/z$ and $\abs{q_*}$.
We assume that the $\cO$ spectrum is measured in twenty equidistant bins between $0 \leq \cO \leq 4 \GeV$
inside 1000 three-dimensional bins in $x, z, Q^2$ with equal statistics $N_{\pi^+}/1000$ in each,
and for definiteness consider a bin centered on $x = 0.1$, $z = 0.15$, $Q = 20 \GeV$ in the following.
To account for the fact that the factorized dependence on $x$, $z$, $Q^2$ is
determined from all bins at equal $x$, $z$, $Q^2$ simultaneously,
we multiply the available statistics by another factor $100$, arriving at
an effective sample size of $N_\mathrm{eff} = N_{\pi^+}/10 = 4.18 \times 10^{7}$.
At fixed $x, z, Q^2$,
a common model for the nonperturbative TMDs is~\cite{Bacchetta:2022awv}
\begin{align} \label{eq:np_model}
\tilde f_1^{\,\NP} &= e^{-\w_1 b_T^2}
\,, \nn \\
\tilde D_1^\NP &= \alpha \,e^{-\w_2 b_T^2} + (1-\alpha) (1 - \w_3 b_T^2) \,e^{-\w_3 b_T^2}
\,,\end{align}
where the $\w_i$ encode the width of the TMDs.
We are interested in how much better the three free parameters $\w_i$
can be determined at the EIC using either $q_*$ or $P_{hT}$.
(We hold the parameter $\alpha$ fixed for simplicity.)
We assume a Gaussian prior probability density $\pi(\w_i)$
based on the central values and standard deviations from \cite{Bacchetta:2022awv}.
We combine \eq{np_model} with leading-logarithmic TMD evolution
and tree-level matching in \texttt{SCETlib}~\cite{scetlib},
and insert it into \eqs{leading_SIDIS_pol}{qstarfact}
to generate EIC pseudodata $d_n$ for bin $n$ in $\cO$, using the central $\w_i$.
In the same way, we generate theory replicas $t_n(\w_i)$ distributed according to the prior.
By normalizing $\sum d_n = \sum t_n = 1$ to the sum over bins at fixed $x$ and $z$,
collinear PDFs and FFs drop out at this order.
(For details on the theory calculation, see \app{reweighting}.)
We then sample the posterior parameter probability distribution
\begin{align}
\pi(\w_i \,|\, d_n) \propto \exp \biggl[- \sum_n\Bigl(\frac{d_n-t_n(\w_i)}{\sigma_n}\Bigr)^2\biggr] \, \pi(\w_i)
\end{align}
using a standard $\chi^2$ likelihood function, where $\sigma_n = \sqrt{d_n / N_\mathrm{eff}}$
is the Poisson error on pseudodata bin $n$.
Our results for the mean and variance of the posterior distribution
are shown in \fig{stat_sensitivity} compared to those of the prior.
Comparing $P_{hT}$ and $q_*$, we find that the superior experimental resolution of $q_*$
only requires giving up a minor amount of statistical sensitivity to the $\w_i$.
In particular, there is
more than a factor 10 improvement in uncertainty on the dominant fragmentation parameter $\w_3$
in either case.
The choice of binning at small $q_*$ should be optimized in the future to exploit its excellent resolution,
but we stress that we have not done so here.

\begin{figure}[t!]
\includegraphics[width=\columnwidth]{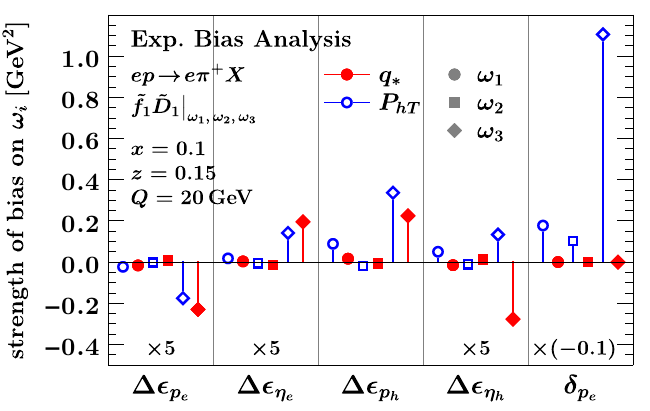}
\vspace{-0.2in}
\caption{%
Strength of bias on TMD nonperturbative model coefficients
from potential sources of systematic uncertainty.
By construction $q_*$ (red) is robust against the large calibration uncertainty $\delta_{p_e}$ that impacts $P_{hT}$ (blue).
Both $q_*$ and $P_{hT}$ exhibit similar susceptibility to non-uniform detector response, modeled by $\Delta \eps_X$.
}
\vspace{-0.2in}
\label{fig:bias_strength}
\end{figure}

The same
setup can be used to assess the robustness of $q_*$
against systematic uncertainties.
We use \texttt{Pythia}~\cite{Bierlich:2022pfr} to generate biased pseudodata $d_n^\mathrm{bias}$ subject
to either (i)~a momentum miscalibration, $p \to (1 + \delta_{p}) \, p$,
or (ii)~a shape effect from a non-uniform detector response (encoded e.g.\ in an efficiency)
that changes at a slow rate $\Delta \eps_X$
as a function of $X = \{p_e, p_h, \eta_e, \eta_h\}$ across the $x, z, Q^2$ bin at hand.
(The absolute value of the efficiency cancels in the normalized $d_n^\mathrm{bias}$.)
Repeating the reweighting analysis, we evaluate the partial derivatives of the posterior's mean $\w_i$
with respect to the bias parameters, which we dub the ``strength'' of the bias, as shown in \fig{bias_strength}.
As anticipated, we find that an analysis using $P_{hT}$ is severely susceptible
to the electron momentum calibration $\delta_{p_e}$,
while the calibration uncertainty using $q_*$ vanishes exactly due to its purely angular nature.
(Both $P_{hT}/z$ and $q_*$ are independent of $\delta_{p_h}$ at observable level for $M_h \ll Q$.)
Figure~\ref{fig:bias_strength} also shows that $P_{hT}$ and $q_*$
have comparable susceptibility to non-uniform detector response,
despite the exponential factors of $\eta_{e,h}$ appearing in $q_*$,
demonstrating the robustness of $q_*$ against these sources of bias.


By replacing measurements of $\df\sigma/\df P_{hT}\df \phi_h$ by $\df\sigma/\df q_*$ with  angular reconstruction of $Q,x,y$, the prospects for precisely mapping the 3D structure of hadronization and confinement with TMDs are bright.
We anticipate a follow-up campaign to aid this endeavor by discovering other useful angular observables
that resolve the remaining TMD PDFs
and by studying theoretical ingredients,
like the convergence of the known higher-order QCD corrections for these cross sections.

\begin{acknowledgments}
\headingAcknowledgments
This work was supported by the U.S.~Department of Energy, Office of Science, Office of Nuclear Physics, from DE-SC0011090.
I.S.~was also supported in part by the Simons Foundation through the Investigator grant 327942.
\end{acknowledgments}

\bibliography{refs}

\onecolumngrid
\clearpage

\makeatletter
\renewcommand\p@subsection{}
\makeatother

\section*{Supplemental material}
\label{app:supplement}

\setcounter{equation}{1000}
\setcounter{figure}{1000}
\setcounter{table}{1000}

\renewcommand{\theequation}{S\the\numexpr\value{equation}-1000\relax}
\renewcommand{\thefigure}{S\the\numexpr\value{figure}-1000\relax}
\renewcommand{\thetable}{S\the\numexpr\value{table}-1000\relax}

\subsection{Constructing $q_*$ with finite target mass}
\label{app:finite_mass}

In the main text, we present our results in the light target mass limit $M \ll Q$.
Here, we give the corresponding results when fully retaining the target mass $M$, which can be important when $Q$ is not sufficiently large or when $M$ is large (such as for a nucleus).  This amounts to including in our construction the dependence on
\begin{align}
\gamma = \frac{2xM}{Q}
\,.\end{align}
We also have the following variable generalizations:
\begin{align}
\eps = \frac{1 - y - \frac{1}{4} y^2 \gamma^2}{1 - y + \frac{1}{4}y^2 \gamma^2 + \frac{1}{2} y^2}
\,, \qquad
\kappa_\gamma = \Bigl[1- \Bigl(\frac{m_{hT} \, \gamma}{z Q} \Bigr)^{2} \Bigr]^{-1/2}
\,, \qquad
m_{hT}^{2} = M_{h}^{2} + P_{hT}^{2}
\,.\end{align}
Note that we still take $P_{hT}, M_{h} \ll Q$ (appropriate for example when $N$ is a proton or ion and $h$ is a pion),
and hence can still take $\kappa_{\gamma} = 1$.
Only $\eps$ receives mass corrections through $\gamma$.
We will show that the above corrections do not change any of the conclusions regarding the utility of $q_*$.

We continue to work with the leading power (LP) kinematics $\la \sim P_{hT}/(zQ) \ll 1$.
For the relationship between the acoplanarity angle $\phi_\acop^\rest$
and the hadron transverse momentum $P_{hT}$, we now have
\begin{align}
\tan\phi_\acop^\rest
&=  \frac{\sin\phi_h \, P_{hT}}{z Q}
\sqrt{\frac{1+4M^2x^2/Q^2}{1-M^2x^2y^2/Q^2-y}}
+ \ord{\la^{2}}
\nn \\
&=  \frac{\sin\phi_h \, P_{hT}}{z Q}
\sqrt{\frac{1+\gamma^{2}}{1-\gamma^{2} y^2 / 4-y}}
+ \ord{\la^{2}}
\,.\end{align}
We now wish to construct an optimized observable $q^{M}_{*}$ for a massive target with $M \sim Q$,
such that $q^{M}_{*} \stackrel{M\ll Q}{=} q_*$ while retaining the desired leading power relation $q^{M}_{*} \stackrel{\text{LP}}{=} -\sin\phiH \frac{P_{hT}}{z}$.

First of all, we emphasize that the following relations presented in the main text
are independent of $M$:
\begin{align}
y = 1- \frac{\sin\theta_{h}}{\cos\alpha} \,, \qquad
Q^{2} = (\ell^{0}_\rest)^{2} \Bigl[\frac{\sin^{2}\theta_{e}}{\cos^{2}\alpha} - \bigl( 1- \frac{\sin\theta_{h}}{\cos\alpha} \bigr)^{2} \Bigr]
\,.\end{align}
Furthermore, we notice that $\gamma^{2}$ can also be written in terms of angles in a manner that is independent of $M$:
\begin{align}
\gamma^{2} = \frac{4 M^{2}x^{2}}{Q^{2}} = \frac{M^{2} Q^{2}}{(P \cdot q)^{2}}
= \Bigl( \frac{\sin\theta_{e}}{\cos\alpha-\sin\theta_{h}} \Bigr)^{2} - 1
\,.\end{align}
This allows us to define $q^{M}_{*}$ in terms of \emph{target rest frame} quantities
completely free of $M$ dependence:
\begin{align} \label{eq:def_q_star_M}
q^{M}_{*} \equiv - \ell^0_\rest \frac{ [\cos(\theta_{e} + \theta_{h}) + \cos\theta_{h} ] \tan(\frac{\theta_{e}}{2}) \sin\theta_{h}}
{\sin(\theta_{e} + \theta_{h})} \, \tan \phi_{\acop} ^\rest
\stackrel{\text{LP}}{=} - \sin \phi_h \frac{P_{hT}}{z}
\,.\end{align}
This relation is especially useful for fixed target experiments,
where all the quantities can be readily measured.
For collider experiments like the EIC, the above equation may be expressed in terms of lab frame quantities
simply by substituting
$\tan\frac{\theta_{i}}{2} = \exp\bigl[\cosh^{-1} (P^0_\EIC/M) + \eta_{i}\bigr]$ for $i = e, \,h$,
where $\eta_{i}$ are the lab frame pseudorapidities,
$\ell^{0}_\rest =  \exp\bigl[\cosh^{-1} (P^0_\EIC/M)\bigr] \ell^0_\EIC$,
and $\phi_\acop^{\rest} = -\phi_{\acop}^{\EIC}$.
We remind the reader that the last relation is due to the different convention
for the orientation of the $z$ axis at the EIC, see \fig{kinematics}.

We emphasize that since $q^{M}_*$ has the desired LP limit $q^{M}_{*} \stackrel{\text{LP}}{=} -\sin\phiH \frac{P_{hT}}{z}$,
the factorization formula for $\df\sigma/(\df x\, \df y \, \df z \, \df q_*^M)$
is the same as written in \eq{leading_SIDIS_pol}, does not receive mass corrections,
and is valid for both fixed-target and collider experiments.

\subsection{Power corrections to double angle formulas for $Q^2$, $y$, and $x$}
\label{app:pc_dbleangle}

In the main text, we give the expression of $Q^{2}$ and $y$ in \eq{dbleangle} in terms of lab frame angles
to leading order in $\la \sim P_{hT}/(zQ)$.
Here, we derive the leading power correction to these kinematic relations.
These results can be used to get an idea of the size of power corrections to an analysis, including both i) the size of corrections to the double angle construction for $x$, $y$, and $Q^2$, and ii) power corrections to the factorization formula for $\df\sigma/(\df x\,\df y\,\df z\,\df q_*)$.
Note that in this section we still work in the massless target limit $M \ll Q$.

We have the following relations:
\begin{subequations} \label{eq:rapidity_energy}
\begin{align}
s &= (P + \ell)^2 = 2 P\cdot \ell = \frac{Q^2}{xy}
\,, \label{eq:rap_Ecm} \\
\frac{(2 P^{0}_\EIC)^{2}} {s}\, \exp(-2\eta_e) &= \frac{P \cdot \ell'}{\ell \cdot \ell'} = \frac{P \cdot q - P \cdot \ell}{\ell \cdot q}
= \frac{1}{x} \frac{1-y}{y}
\,, \label{eq:rap_etae}\\
\frac{(2 P^{0}_\EIC)^{2}} {s}\exp(-2\eta_h) &= \frac{P \cdot P_h}{\ell \cdot P_h}
= \frac{P \cdot q}{x (\ell \cdot P) + \ell \cdot q +\frac{1}{z} \, \ell\cdot P_{hT}}
\nn \\
&= \frac{y}{x (1-y)}  \Bigl(1+ \frac{2 \cos\phiH}{\sqrt{1-y}} \, \frac{P_{hT}}{zQ} \Bigr)
+ \ord{\la^{2}}
\,, \label{eq:rap_etah} \end{align}
\end{subequations}
where $\eta_{e}$ and $\eta_{h}$ are lab frame pseudorapidities.
Note that we have indicated that \eq{rap_etah} is the only equation here that receives corrections in $\la\sim
P_{hT}/(zQ)$ when expanding in this ratio.
Solving the above relations for $x$, $y$, and $Q^{2}$,
we get
\begin{subequations}\label{eq:la_correction}
\begin{align}
x &=  \frac{(2 P^{0}_\EIC)^{2}} {s} e^{\eta_{e}+ \eta_{h}} \, \biggl(1 + \cos\phiH \sqrt{1+ e^{-\Delta\eta}} \, \frac{P_{hT}}{zQ} \biggr)
+ \ord{\la^{2}},
\\
y &= \frac{1}{1+e^{\Delta\eta}} + \ord{\la^{2}},
\label{eq:la_correction_y}
\\
Q^{2} &= (2 P^{0}_\EIC)^{2} \frac{ e^{\eta_{e} + \eta_{h}} }{1+e^{\Delta\eta} } \, \biggl(1 + \cos\phiH \sqrt{1+ e^{-\Delta\eta}} \, \frac{P_{hT}}{zQ}\biggr)
+ \ord{\la^{2}}
\, ,\end{align}
\end{subequations}
where $\Delta\eta = \eta_{h}-\eta_{e}$ is boost invariant along the $z$-direction.
Notice that in \eq{la_correction_y}, $y$ does not receive linear corrections in $\la$.
An application of \eq{la_correction} is to test the size of the power corrections in the expressions for $x$ and $Q^2$ in a given data set, and thus apply cuts to restrict the data to TMD regions where the leading term is dominant.

We can also invert the formulae in \eq{la_correction} to define a set of variables $x_{*}, \, y_{*}$ and $Q_{*}$ that use lab frame measurements.
The variables $x_{*}, \, y_{*}$ and $Q_{*}$ agree with the kinematic invariants $x,\, y$ and $Q$ up to the determined ${\cal O}(\lambda)$  power corrections:
\begin{subequations}\label{eq:la_correction_inverted}
\begin{align}
x_{*} &\equiv \frac{\bigl(2  P_\EIC^0 \bigr)^2}{s} e^{\eta_{e} + \eta_{h}} \, = x \biggl(1 - \cos\phiH \sqrt{\frac{1}{1-y}} \, \frac{P_{hT}}{zQ} \biggr)
+ \ord{\la^{2}},
\\
y_{*} &\equiv \frac{1}{1+e^{\Delta\eta}} = y
+ \ord{\la^{2}},
\\
Q_{*}^{2} &\equiv \bigl(2 P_\EIC^0 \bigr)^{2} \frac{ e^{\eta_{e} + \eta_{h}} }{1+e^{\Delta\eta} } \, = Q^{2}  \biggl(1 - \cos\phiH  \sqrt{\frac{1}{1-y}} \, \frac{P_{hT}}{zQ} \biggr)
+ \ord{\la^{2}}
\,.\end{align}
\end{subequations}
The all-order definition of $q_{*}$ in the main text, \eq{qstar}, can be written as
$q_{*} \equiv Q_{*} \sqrt{1-y_{*}} \tan\phi_{\acop}^\EIC$.
This allows us to easily determine the leading power correction to the formula for $q_{*}$
obtained by expanding in $P_{hT}\ll zQ$:
\begin{align}
q_{*} = - \sin\phiH \frac{P_{hT}}{z} \biggl( 1-\frac{\cos\phiH}{2}
\sqrt{\frac{1}{1-y}} \, \frac{P_{hT}}{z Q} \biggr)
+ \ord{\la^{3}}
\, .\end{align}
This kinematic correction to the relationship between variables is the only power correction that would give non-trivial dependence on $y$ and $\cos\phi_h$
to the factorization formula in \eq{qstarfact}. Hence it can be unambiguously included in the factorization analysis by using this more complicated relationship in the $\delta(q_* + \ldots)$ when switching variables and integrating over $P_{hT}$ and $\phi_h$ (cf.\ the example given in \eq{FactCos2phih} without these corrections).  However, this still does not capture the dynamic hadronic power corrections, which arise from using the $P_{hT}\ll zQ$ expansion when deriving the original factorization theorem for $\df\sigma/(\df x\,\df y\,\df z\,\df^2\vec P_{hT})$.

\subsection{Leading-power formulas for target spin vector from angular measurements}
\label{app:angular_target_spin}

As mentioned in the main text,
the $S_L$, $S_T$ and $\phiS$ that appear in the factorization formula are defined in the Trento frame
by writing the  nucleon spin vector as $S^\mu = (0, S_T \cos \phi_S, S_T \sin \phi_S, -S_L)_\mathrm{Trento}$.
Here we give leading-power expressions for these variables in terms of the target rest frame components of $S^\mu=(0,S_x,S_y,S_z)_\rest$,
and the EIC lab frame hadron pseudorapidity $\eta_h$.
Note that we do not assume that the nucleon is in a pure spin state.

We start with expressing $S_L$, $S_T$ and $\phiS$ using the polar angle $\theta_q$ of $\vec q$,
\begin{align}
S_L=-S_z \cos\theta_q + S_x\sin\theta_q\,,\qquad S_T = \sqrt{(S_x \cos\theta_q+S_z\sin\theta_q)^2+S_y^2}\,.
\end{align}
$\phiS$ is given by
\begin{align}
\sin\phiS = \frac{S_y}{S_T}\,, \qquad \cos\phiS = \frac{S_x \cos\theta_q+S_z\sin\theta_q}{S_T}\,.
\end{align}
At leading power in $\lambda\sim P_{hT}/(zQ)\ll 1$,  we may replace $\theta_q$ by $\theta_h$,
which can be written in terms of $\eta_h$.
We have
\begin{align}
S_L&=-S_z\, \frac{1-A^2\, e^{2\eta_h}}{1+A^2\, e^{2\eta_h}}+ S_x\, \frac{2A\, e^{\eta_h}}{1+A^2\, e^{2\eta_h}} +\Ord{\lambda}\,,\\[.1in]
S_T &=\sqrt{ \left[\frac{(1-A^2\, e^{2\eta_h})S_x +2A\, e^{\eta_h} S_z }{1+A^2\, e^{2\eta_h}}\right]^2+S_y^2 } \, +\Ord{\lambda}
\,,\nn\\[.1in]
\sin\phiS &= \frac{S_y}{S_T}\,, \qquad \cos\phiS = \frac{(1-A^2\, e^{2\eta_h})S_x +2A\, e^{\eta_h} S_z }{(1+A^2\, e^{2\eta_h})S_T} +\Ord{\lambda}\,.
\end{align}
Here we define $A$ to be
\begin{align}
A=\exp\bigl[\cosh^{-1} (P^0_\EIC/M)] \stackrel{M \ll P^0_\EIC}{\approx} \frac{2P^0_\EIC}{M}\,.
\end{align}

\subsection{Momentum space factorization formula for the $q_*$ spectrum}
\label{app:momentum_space_factorization}

In the main, text our factorization theorem for $\df\sigma/(\df x\,\df y\,\df z\,\df q_*)$  was written in terms of $b_T$ space TMDs.
For completeness, here we give the factorization theorem written in terms of momentum space TMDs, which are Fourier conjugate to those in $b_T$ space.

We start with the standard leading power momentum space TMD factorization formulae~\cite{Bacchetta:2006tn} for the structure functions appearing in \eq{leading_SIDIS_pol}:
\begin{align} \label{eq:leadingWmom}
W_{UU,T} &=
\tilde \cF \left[ \cH^{(0)}\, f_1 D_1\right]
\,,\nn\\
W_{UU}^{\cos 2\phi_h} &=
\tilde\cF \left[\frac{-2 (\hat h\cdot\kt) (\hat h\cdot\pt) + \kt \cdot \pt}{M M_h} \cH^{(0)}\, h_1^{\perp} H_1^{\perp}\right]
\,,\nn\\
W_{UL}^{\sin2\phi_h} &=
\tilde\cF \left[\frac{-2 (\hat h\cdot\kt) (\hat h\cdot\pt) + \kt \cdot \pt}{M M_h} \cH^{(0)}\, h_{1L}^{\perp}H_1^{\perp}\right]
\,,\nn\\
W_{LL} &=
\tilde\cF \left[\cH^{(0)}\, g_{1L} D_1\right]
\,,\nn\\
W_{UT,T}^{\sin\left(\phi_h-\phi_S\right)} &=
\tilde\cF \left[ -\frac{\hat h\cdot\kt}{M} \cH^{(0)}\, f_{1T}^{\perp} D_1\right]
\,,\nn\\
W_{UT}^{\sin\left(\phi_h+\phi_S\right)} &=
\tilde\cF \left[ -\frac{\hat h\cdot\pt}{M_h} \cH^{(0)}\, h_1 H_1^{\perp}\right]
\,,\nn\\
W_{UT}^{\sin\left(3\phi_h-\phi_S\right)} &=  \tilde\cF \left[\frac{2 (\hat h\cdot\kt)\, (\vec k_T\cdot \vec p_T) + k_T^2\, (\hat h\cdot\pt) - 4(\hat h\cdot\kt)^2\, (\hat h\cdot\pt)}{2 M^2M_h}\, \cH^{(0)} h_{1T}^\perp H_1^\perp\right]
\,,\nn\\
W_{LT}^{\cos\left(\phi_h-\phi_S\right)} &= \tilde\cF \left[ \frac{\hat h\cdot\kt}{M} \cH^{(0)}\, g_{1T} D_1\right]
\,,\end{align}
where $\tilde\cF$ is defined as
\begin{align}\label{eq:cF_tilde}
\tilde\cF[\w\, \cH\, g\, D] &
\equiv 2z \sum_f \cH_f(Q^2) \int\!\df^2\kt\, \df^2\pt \, \delta^{(2)}\bigl(\qt+\kt-\pt\bigr)
\times \w(\vec k_T, \vec p_T)\, g_f(x, k_T) \, D_f(z,p_T)
\,.\end{align}
Here $\vec q_T=-\vec P_{hT}/z$, $\hat h=\vec P_{hT}/P_{hT}$, and
$\w(\kt,\pt)$ denotes the weight prefactors in \eq{leadingWmom} that depend on $\kt$ and $\pt$.

The leading-power SIDIS cross section differential in $q_*$ is
\begin{align}\label{eq:dsigma_dqstar}
& \frac{\df \sigma}{\df x \, \df y \, \df z \, \df q_*}
= \int_0^\infty \! \df P_{hT} \, P_{hT} \int_0^{2\pi} \! \df \phiH \, \delta \bigl( q_* + \sin \phi_h P_{hT}/z \bigr)
\frac{\df\sigma}{\df x \, \df y \, \df z \, \df^2\PTb}
\,.\end{align}
Plugging \eq{leading_SIDIS_pol} into \eq{dsigma_dqstar}, since the structure functions themselves have no dependence on $\phiH$, terms which have angular prefactors that are odd under $\phiH \to \pi-\phiH~$ vanish under the integral of $\df\phi_h$.
Thus we are left with
\begin{align}\label{eq:dsigma_dqstar_terms}
\frac{\df \sigma}{\df x \, \df y \, \df z \, \df q_*} &= \sigma_0 \int_0^\infty \! \df P_{hT} \, P_{hT} \int_0^{2\pi} \! \df \phiH \, \delta \bigl( q_* + \sin \phi_h P_{hT}/z \bigr)
\Bigl\{
W_{UU,T} + \lambda_e S_L \sqrt{1-\eps^2} \, W_{LL}
\nn \\ & \:\;
+ \eps \cos(2\phiH) W_{UU}^{\cos(2\phiH)}
+ S_T \sin\phiH \cos\phiS \left( W_{UT,T}^{\sin(\phiH \!- \phiS)}
+ \eps W_{UT}^{\sin(\phiH \!+ \phiS)}\right)
\nn \\& \:\;
+ S_T \,  \cos\phiS \sin (3 \phiH) W_{UT}^{\sin(3 \phiH \! -\phiS)}
+ \lambda_e S_T \sqrt{1-\eps^2} \sin\phiH \sin\phiS W_{LT}^{\cos(\phiH \! -\phiS)}
\Bigr\}
\,.\end{align}
Especially notice that $W_{UL}^{\sin2\phiH}\sim h_{1L}^\perp H_1^\perp$ does not appear in \eq{dsigma_dqstar_terms} since its prefactor $\sin(2\phi_h)$ is odd under $\phi_h\to \pi-\phi_h$.

For each term in \eq{dsigma_dqstar}, writing the $\phi_h$ dependent coefficient as $\kappa(\phi_h)$, we have
\begin{align}\label{eq:dsig_dqstar_intqT}
\frac{\df \sigma}{\df x \, \df y \, \df z \, \df q_*} [\kappa(\phi_h)\, W (\w \cH g D) ] &=  2z \sum_f \cH_f(Q^2) \int\!\df^2 \vec P_{hT}\, \df^2\kt\, \df^2\pt \, \delta^{(2)}\bigl(\qt+\kt-\pt\bigr) \, \delta\bigl( q_* + \sin \phi_h P_{hT}/z \bigr)
\nn\\ &\hspace{1in}
\times \kappa(\phi_h)\, \w(\vec k_T, \vec p_T)\, g_f(x, k_T) \, D_f(z,p_T)
\nn\\[.1in] &=
2z^{3} \sum_f \cH_f(Q^2) \int\! \df^2\kt\, \df^2\pt \, \delta\bigl( q_* + \hat y\cdot (\kt-\pt)\bigr)
\nn\\ &\hspace{1in}
\times \kappa(\kt,\pt)\, \w(\vec k_T, \vec p_T)\, g_f(x, k_T) \, D_f(z,p_T)
\,,\end{align}
where $\hat y$ is the unit vector in the Trento frame.
In the second line, all appearances of $\hat h$ in $\w(\kt,\pt)$ are replaced by
\begin{align}
\hat h = \frac{\kt-\pt}{\abs{\kt-\pt}}\,,
\end{align}
and $\kappa(\phi_h)$ is replaced by a function of $\kt$ and $\pt$ according to the rule
\begin{align}
\cos\phi_h \to \hat h \cdot \hat x'\,, \qquad
\sin\phi_h \to \hat h \cdot \hat y
\,,\end{align}
where $\hat x'$ and $\hat y$ are again Trento frame unit vectors.

Then we can write \eq{dsigma_dqstar_terms} as
\begin{align}
& \frac{1}{\sigma_0} \frac{\df \sigma}{\df x \, \df y \, \df z \, \df q_*} =
\tilde\cF_*[f_1 D_1]
+ \lambda_e S_L \sqrt{1-\eps^2} \,\tilde\cF_*[g_{1L} D_1]
\nn \\[.1in] & \qquad
+ \eps\, \tilde\cF_*\!\left[ \left[(\hat h \cdot \hat x')^2 - (\hat h \cdot \hat y)^2\right]\frac{ -2 (\hat h\cdot\kt) (\hat h\cdot\pt) + \kt \cdot \pt}{M M_h}\, h_1^{\perp} H_1^{\perp}\right]
\nn \\[.1in] & \qquad
+ S_T \cos\phiS \Biggl(  \tilde\cF_*\left[(\hat h \cdot \hat y) \frac{-\hat h\cdot\kt}{M} f_{1T}^{\perp} D_1\right]
+ \eps \, \tilde\cF_*\!\left[(\hat h \cdot \hat y) \frac{-\hat h\cdot\pt}{M_h} h_1 H_1^{\perp}\right] \Biggr)
\nn \\[.1in]& \qquad
+ S_T \cos\phiS\, \tilde\cF_*\!\left[\left[ 3 (\hat h \cdot \hat y) - 4 (\hat h \cdot \hat y)^3 \right] \frac{2 (\hat h\cdot\kt)\, (\vec k_T\cdot \vec p_T) + k_T^2\, (\hat h\cdot\pt) - 4(\hat h\cdot\kt)^2\, (\hat h\cdot\pt)}{2 M^2M_h}\, h_{1T}^\perp H_1^\perp\right]
\nn \\[.1in]& \qquad
+ \lambda_e S_T \sqrt{1-\eps^2} \sin\phiS\,  \tilde\cF_*\!\left[(\hat h \cdot \hat y) \frac{\hat h\cdot\kt}{M} g_{1T} D_1\right]
\,.\end{align}
Note that we have used $\sin (3 \phiH) = 3 \sin\phiH - 4 \sin^{3}\phiH$.
Here we define $\tilde\cF_*$ as
\begin{align}
\tilde\cF_*[\w_*\, \cH\, g\, D] \equiv 2 z^{3} \sum_f \cH_f(Q^2)\int\! \df^2\kt\, \df^2\pt \, \delta\bigl( q_* + \hat y\cdot \kt-\hat y\cdot\pt\bigr)\,
\w_*(\vec k_T, \vec p_T)\, g_f(x, k_T) \, D_f(z,p_T)\,,
\end{align}
where the weight function $\w_*(\kt,\pt)$ includes the full prefactor structures, and can be understood as product of $\kappa(\kt,\pt)$ and $\w(\kt,\pt)$ in \eq{dsig_dqstar_intqT}.

\subsection{Resolution curves for all detector regions}
\label{app:resolution_curves}

\begin{figure}[t!]
\includegraphics[width=0.49\columnwidth]{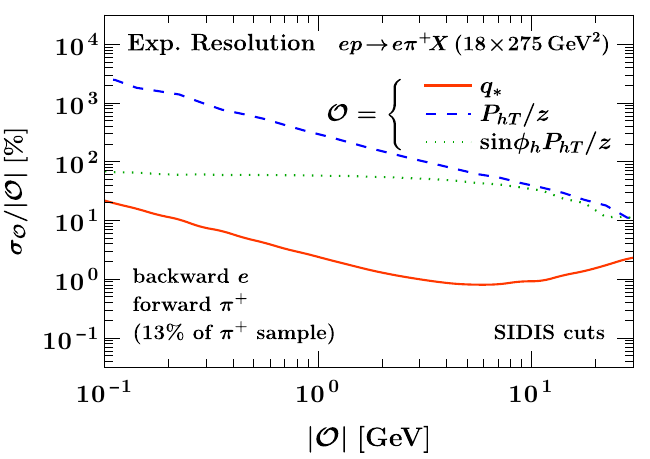}
\includegraphics[width=0.49\columnwidth]{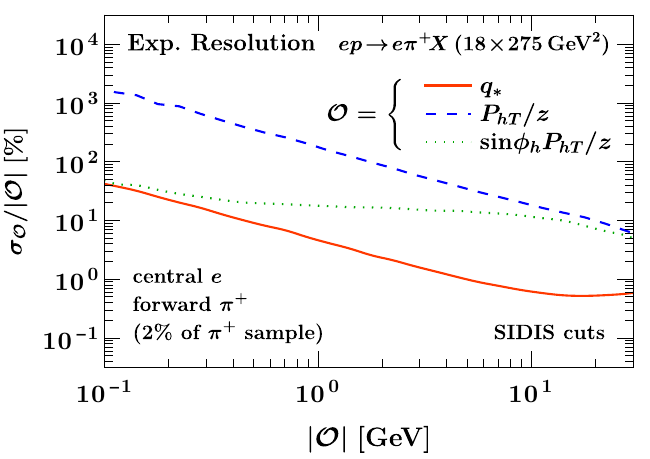}
\\
\includegraphics[width=0.49\columnwidth]{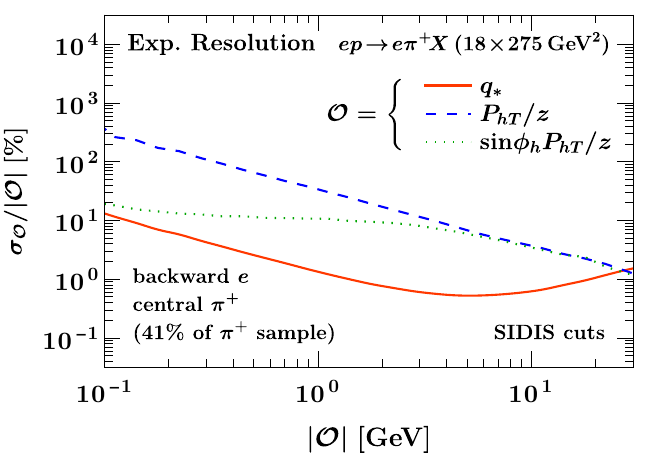}
\includegraphics[width=0.49\columnwidth]{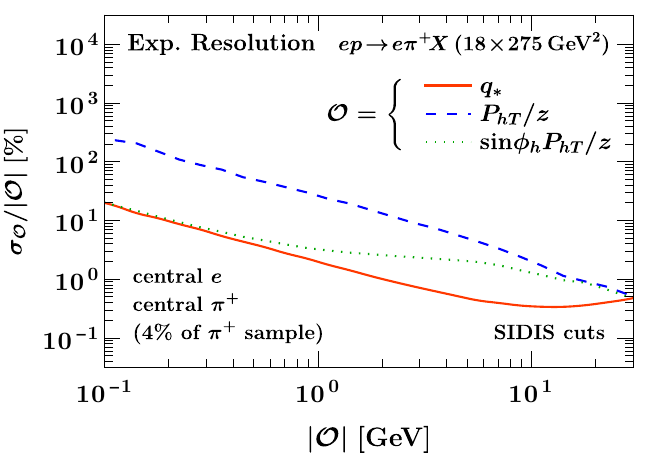}
\\
\includegraphics[width=0.49\columnwidth]{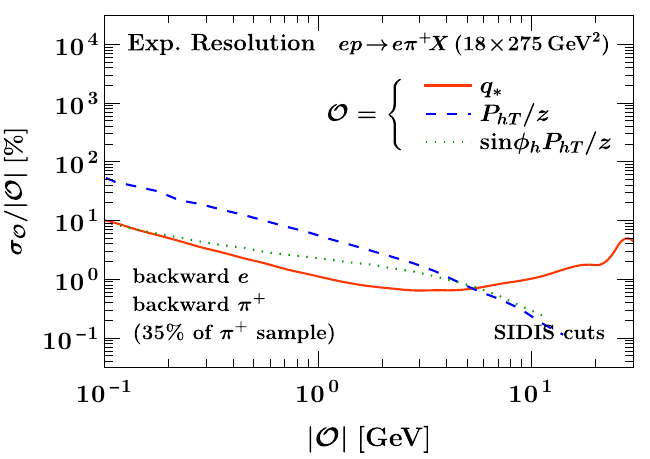}
\includegraphics[width=0.49\columnwidth]{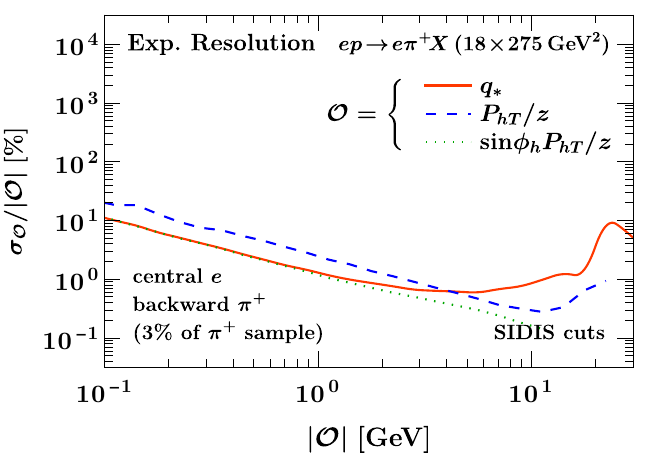}
\caption{%
Expected event-level detector resolution $\sigma_\cO$ for different SIDIS TMD observables $\cO$.
We show relative resolutions as a function of the magnitude of $\cO = q_*$ (solid red), $P_{hT}/z$ (dashed blue), and $P_{hT}/z \sin \phi$ (dotted green),
for backward electrons (left column) and central electrons (right column)
as well as forward pions in the top row, central pions in the center row, and  backward pions in the bottom row.
The center left panel corresponds to \fig{resolution_curves_bc} in the main manuscript.
}
\vspace{-0.2in}
\label{fig:resolution_curves_all}
\end{figure}

\fig{resolution_curves_bc} in the main manuscript was restricted to the case
of a backward electron and central pion, which features the largest share
of the total pion sample for our selection cuts.
In \fig{resolution_curves_all}, we provide additional results
for the expected detector resolution of different SIDIS TMD observables
in all other relevant detector regions.
Note the change in vertical scale compared to \fig{resolution_curves_bc}.
We find that a clear improvement in resolution from using $q_*$ persists
across all detector regions.
The case of electrons in the forward detector region
(i.e., from backscattered electrons at very large $Q^2 \to s$)
has a negligible contribution to the total rate.
We note that the improvement of $q_*$ in resolution compared to $P_{hT}$
deteriorates slightly in the cases where the hadron is backward for $P_{hT}< 5\,{\rm GeV}$,
and actually features worse resolution for $P_{hT} > 5\,{\rm GeV}$.
This is expected as the fixed angular resolution $\sigma_{\theta_h^\EIC} = 0.001$
translates to a wider range in pseudorapidity as $\theta_h^\EIC \to \pi$.

\subsection{Details on Bayesian reweighting}
\label{app:reweighting}

\subsubsection{Theory templates}

In this section we describe in detail how the pseudodata $d_i$ and theory replicas $t_i(\w_i)$
in the main text are constructed. They are defined as the normalized bin-integrated spectrum
for the observable $\cO = P_{hT}/z$ or $\cO = q_*$ being tested,
\begin{align} \label{eq:def_template}
\{d_n, t_n(\w_i)\} = \frac{1}{\df \sigma/(\df x \, \df z \, \df Q^2)} \int_{\cO_{a,n}}^{\cO_{b,n}} \! \df \cO \, \frac{\df \sigma}{\df x \, \df z \, \df Q^2 \, \df \cO}
\,,\end{align}
where the right-hand side is evaluated with the appropriate
values of $\w_i$ (either central or chosen according to the Monte-Carlo replica) inserted into \eq{np_model}.
Restricting \eqs{leading_SIDIS_pol}{qstarfact}
to the unpolarized contribution from $W_{UU,T}$, the differential leading-power spectrum
explicitly reads
\begin{align}
\frac{\df \sigma}{\df x \, \df z \, \df Q^2 \, \df \cO}
= \frac{2yz^3}{Q^2} \, \sigma_0 \int_0^\infty \! \df b_T \, K_\cO(\cO b_T)
\sum_f \cH_f(Q^2, \mu) \, \tilde f_{1f}(x, b_T, \mu, \zeta) \, \tilde D_{1f}(z, b_T, \mu, \zeta)
\,,\end{align}
where $\mu$ is the \MSbar{} scale that the TMD PDF and FF are being evolved to,
$\zeta = Q^2$ is the Collins-Soper scale, and
\begin{align}
K_{P_{hT}/z}(x) = x J_0(x)
\,, \qquad
K_{q_*} = \frac{2 \cos(x)}{\pi} \,,
\end{align}
is the integral kernel that depends on the respective observable.
At tree-level, the hard function simply reduces to the electric charge, $\cH_f = e_f^2$.
Using tree-level matching onto collinear PDFs and FFs,
the evolved TMDs are given by
\begin{align}
\tilde f_{1f}(x, b_T, \mu, \zeta)
&= f_{1f}(x, \mu) \, U(b_T, \mu_0, \zeta_0, \mu, \zeta) \, \tilde f_1^{\,\NP}(x, b_T)
\,, \nn \\
\tilde D_{1f}(z, b_T, \mu, \zeta)
&= D_{1f}(z, \mu) \, U(b_T, \mu_0, \zeta_0, \mu, \zeta) \, \tilde D_1^\NP(z, b_T)
\,,\end{align}
where $\tilde f_1^{\,\NP}$ and $\tilde D_1^\NP$ are the nonperturbative model function of interest
and satisfy $\tilde f_1^{\,\NP}(x, b_T), \tilde D _1^\NP(z, b_T) = 1 + \ord{\lqcd^2 b_T^2}$.
Here we have used that at our leading-logarithmic working order,
we are free to evaluate the collinear PDF $f_{1f}(x, \mu)$ and collinear FF $D_{1f}(z, \mu)$
at the high scale $\mu \sim Q$.
The evolution factor $U$ accounts for the virtuality and Collins-Soper evolution
of the TMD PDF and at leading-logarithmic order is given by
\begin{align} \label{eq:evol_factor}
U(b_T, \mu_0, \zeta_0, \mu, \zeta)
= \exp \Bigl[
- \int_{\mu_0}^{\mu} \! \frac{\df \mu'}{\mu'}
\frac{\alphas(\mu')}{2\pi} \,
\Gamma_0^q \ln \frac{\sqrt{\zeta}}{\mu'}
\Bigr]
\,\exp \Bigl[ \frac{1}{2} \tilde \gamma_\zeta^\NP(b_T) \ln \frac{\zeta}{\zeta_0} \Bigr]
\,,
\end{align}
where $\Gamma_0^q = 4C_F$ is the one-loop quark cusp anomalous dimension
and $\tilde \gamma_\zeta^\NP(b_T) = \ord{\lqcd^2 b_T^2}$ is the nonperturbative contribution
to the Collins-Soper kernel.
The initial scales $\mu_0$ and $\zeta_0$ define
the boundary condition at which the nonperturbative TMDs are defined.
We use the prescription~\cite{Bacchetta:2019sam, Bacchetta:2022awv}
\begin{align}
\mu_0(b_T) = \frac{b_0}{b_\mathrm{max}} \Bigl(
\frac{1 - e^{-b_T^4/b_\mathrm{max}^4}}{1-e^{-b_T^4/b_\mathrm{min}^4}}
\Bigr)^{-1/4}
\,, \quad
b_0 = 2 e^{\gamma_E} \approx 1.123
\,, \quad
b_\mathrm{max} = b_0 \GeV^{-1}
\,, \quad
b_\mathrm{min} = \frac{b_0}{\mu}
\,, \quad
\zeta_0 = 1 \GeV^2
\,.\end{align}
This ensures that the coupling $\alphas(\mu')$ under the integral in \eq{evol_factor}
remains perturbative due to $\mu' \geq b_0/b_\mathrm{max} = 1 \GeV$.
In addition, $\mu_0(b_T \to 0) \to b_0/b_\mathrm{min} = \mu$
together with $\gamma_\zeta^\NP \to 0$
ensures that $U \to 1$ for $b_T \to 0$.
We use N$^3$LL $\alphas$ evolution with $n_f = 5$ active flavors
and evolve to $\mu = \sqrt{\zeta} = Q$ throughout our predictions.

Several further simplifications apply to \eq{def_template}.
We may analytically perform the bin integral
over $\cO_a \leq \cO \leq \cO_b$,
\begin{align}
\tilde{K}_\cO(\cO_b b_T) - \tilde{K}_\cO(\cO_a b_T)
\equiv b_T \int_{\cO_a}^{\cO_b} \! \df \cO \, K_\cO(\cO b_T)
\,, \qquad
\tilde{K}_{P_{hT}/z}(x) = x J_1(x)
\,, \qquad
\tilde{K}_{q_*}(x) = \frac{2 \sin(x)}{\pi}
\,.\end{align}
An important limit is the total integral over $\cO$, which involves the integral kernel
\begin{align} \label{eq:limit_total_integral}
\lim_{\cO_b \to \infty} \frac{1}{b_T} \tilde{K}_{P_{hT}/z}(\cO_b b_T )
= \lim_{\cO_b \to \infty} \frac{1}{b_T} \tilde{K}_{q_*}(\cO_b b_T ) = \delta(b_T)
\,.\end{align}
This allows us to analytically evaluate the total cross section
that appears in the denominator in \eq{def_template},
\begin{align}
\frac{\df \sigma}{\df x \, \df z \, \df Q^2}
= \frac{2yz^3}{ Q^2} \, \sigma_0
\sum_f e_f^2 \, f_{1f}(x, \mu) \, D_{1f}(z, \mu)
\,,\end{align}
which recovers the tree-level SIDIS total cross section.
Taking the ratio in \eq{def_template}, many factors drop out,
including in particular the sum over collinear PDFs and FFs,
leaving behind only the flavor-independent model functions and TMD evolution,
\begin{align} \label{eq:final_result_template}
\{d_n, t_n(\w_i)\} = \int_0^\infty \! \frac{\df b_T}{b_T} \,  \Bigl[
\tilde{K}_{\cO}(b_T \cO_{b,n}) - \tilde{K}_{\cO}(b_T \cO_{a,n})
\Bigr] \,
\bigl[ U(b_T, \mu_0, \zeta_0, \mu, \zeta) \bigr]^2 \, \tilde f_1^{\,\NP}(x, b_T) \, \tilde D_1^\NP(z, b_T)
\,,\end{align}
for $\cO = P_{hT}/z$ and $\cO = q_*$.
It is easy to see from \eq{final_result_template} and \eq{limit_total_integral}
that the pseudodata and theory templates indeed satisfy
$\sum_n d_n = \sum_n t_n(\w_i) = 1$,
where we note that we include an overflow bin $\cO \geq 4 \GeV$
in all sums over $n$, including in particular the likelihood function used in the main text.

The simple form of the normalized $P_{hT}/z$ and $q_*$ spectrum in \eq{final_result_template}
in terms of the underlying nonperturbative TMD distributions
ensures that our conclusions about the TMD sensitivity
are fully general, even though we consider fixed $x$ and $z$.
This is because $x$ and $z$ only enter through $\tilde f_1^{\,\NP}(x, b_T)$ and $\tilde D_1^\NP(z, b_T)$,
which for fixed $x$ and $z$ map onto specific values of $\w_{\{1,2,3\}}$ and $\alpha$
in \eq{np_model} as discussed below,
and hence only affect the initial values of our study.
We stress again that the analytic relation
we exploited between the cross sections in TMD and collinear factorization
is due to our leading-logarithmic working order and the specific perturbative scheme choices we made,
but the simplifications above are fully justified for an exploratory study like this one.
In particular, \eq{final_result_template} still retains all the key characteristic
features of a realistic TMD study such as the correct amount of Sudakov suppression
at long distances.

\subsubsection{Nonperturbative model parameters and Bayesian priors}

We work with the nonperturbative model used in the MAPTMD22 global fit
of TMD PDFs and FFs from Drell-Yan and SIDIS data~\cite{Bacchetta:2022awv}.
The model for the CS kernel reads
\begin{align}
\tilde \gamma_\zeta^\NP(b_T) = g_K(b_T^2) = -g_2^2 \frac{b_T^2}{2}
\end{align}
Since our reweighting study is performed at fixed $Q$,
we do not expect to be sensitive to CS evolution between different values of $Q$
and hold the parameter $g_2 = (0.248 \pm 0.008) \GeV$ fixed at its central value.
(A full four-dimensional measurement differential in $x$, $z$, $Q$, and $q_*$
would of course exhibit the usual sensitivity to the CS kernel.)

For fixed $x$ and $z$, the nonperturbative models of \cite{Bacchetta:2022awv} for the TMD PDF and FF relate to \eq{np_model} as
\begin{align}
\w_1 = \frac{g_1(x)}{4}
\,, \qquad
\w_2 = \frac{g_3(z)}{4z^2}
\,, \qquad
\w_3 = \frac{g_{3B}(z)}{4z^2}
\,, \qquad
\alpha = \frac{g_3(z)}{g_3(z) + \lambda_F \, g_{3B}^2(z)/z^2}
\,,\end{align}
where $g_{\{1, 3, 3B\}}$ are functions of $x$ and $z$ defined
in terms of the underlying model parameters in \cite{Bacchetta:2022awv}.
In the comparison, we have used that the fit result of \cite{Bacchetta:2022awv}
down to $x \geq 0.1$ is compatible with a single Gaussian for the TMD PDF
to a good approximation, and have set $\lambda = \lambda_2 = 0$.
For $x = 0.1$ and $z = 0.15$ as considered in the main text,
the first three parameters evaluate to
\begin{align}
\w_1 = (0.0791 \pm 0.0063) \GeV^2
\,, \qquad
\w_2 = (0.0167 \pm 0.0059) \GeV^2
\,, \qquad
\w_3 = (0.8153 \pm 0.0637) \GeV^2
\,,\end{align}
In our reweighting study,
we take the variance of the Gaussian prior probability distributions for $\{ \w_1, \w_2, \w_3 \}$
to be numerically equal to the $1\sigma$ confidence intervals above.
(We ignore nondiagonal entries in the covariance matrix of \cite{Bacchetta:2022awv},
which were found to be small there.)
We hold $\alpha = 0.079 \pm 0.029$ fixed at its central value for simplicity.
We note that we have also performed the reweighting study
using a simplified version of the nonperturbative model used in \cite{Scimemi:2019cmh},
which features an approximately exponential term at large distances,
and have arrived at similar conclusions.

\subsubsection{Parametrizing systematic bias}

\begin{table}[t]
\renewcommand{\arraystretch}{1.5}
\tabcolsep 5pt
\centering
\begin{tabular}{c||c|c|c|c}
\hline\hline
$X$ &  $p_e/\!\GeV$ & $\eta_e$ & $p_h/\!\GeV$ & $\eta_h$
\\ \hline
$\langle X \rangle$ & 19.86& -0.472& 4.23& 0.892
\\[0.5ex]
$\Delta X$ & 0.32& 0.011& 2.18& 0.275
\\[0.5ex]\hline\hline
\end{tabular}
\caption{%
Mean and variance of kinematic distributions in $X = \{ p_e, p_h, \eta_e, \eta_h \}$
for a bin with $0.085 \leq x \leq 0.115, 0.13 \leq z \leq 0.17, 400 \GeV^2\leq Q^2 \leq 401\GeV^2$
as relevant for the biased event weights in \eq{eps_X}.
}
\label{tab:mean_variance_X}
\end{table}

Here we describe how the bias strengths shown in \fig{bias_strength}
in the main text are evaluated.
To model the effect of a momentum miscalibration $\delta_p$,
we use the \texttt{Pythia} sample of pions described in the main text,
restrict to a bin $0.085 \leq x \leq 0.115, 0.13 \leq z \leq 0.17, 400 \GeV^2\leq Q^2 \leq 401\GeV^2$
centered on our choices for $x, z, Q^2$ in the main text,
and evaluate the $\cO = P_{hT}/z$ or $\cO = q_*$ spectrum after replacing
\begin{align}
p_e \to (1 + \delta_{p_e}) \, p_e
\end{align}
in the event record.
Note that we only use the biased event record to calculate $\cO$,
but continue to cut on the true values of $x$, $z$, $Q^2$
since we expect the impact of $\delta_{p_e}$ on the reference Born kinematics
to be subleading compared to the direct effect on the reconstructed $\cO$.
Similarly, to model the effect of a (generic) non-uniform detector response
that changes as a function of $X = \{ p_e, p_h, \eta_e, \eta_h \}$,
we apply an additional weight $\eps(X)$ to each event calculated as
\begin{align} \label{eq:eps_X}
\eps(X) = 1 + \Delta \eps_X \frac{
X - \langle X \rangle
}{
\Delta X
}
\,, \qquad
\Delta X = \sqrt{\langle X^2 \rangle - \langle X \rangle^2}
\,,\end{align}
where the $\langle \dots \rangle$ refers to an average over all events
in the current $x, z, Q^2$ bin.
Normalizing the effect of $\Delta \eps_X$ to the variance of the distribution
ensures that the impacts of $\Delta \eps_X$ for different $X$ are comparable
even when individual $X$ have more or less narrow underlying distributions
or different mass dimension.
The explicit values of $\langle X \rangle$, $\Delta X$ for the bin we consider
are collected in \tab{mean_variance_X} for reference.

To obtain the final biased theory templates,
we take one of the bias parameters
$B = \{
\Delta \eps_{p_e},
\Delta \eps_{\eta_e},
\Delta \eps_{p_h},
\Delta \eps_{\eta_h},
\delta_{p_e}
\}$ to be nonzero at a time and evaluate the biased normalized pseudodata as
\begin{align} \label{eq:pseudodata_biased}
d_n^\mathrm{bias} = d_n \frac{h_n^\mathrm{bias}}{h_n} \biggl( \sum_m d_m \frac{h_m^\mathrm{bias}}{h_m} \biggr)^{-1}
\,,\end{align}
where $h_n$ ($h_n^\mathrm{bias}$) is the \texttt{Pythia} result
for the normalized $\cO$ spectrum in the same bins with vanishing (nonzero) bias.
We then insert the biased pseudodata from \eq{pseudodata_biased}
into the reweighting analysis, using unbiased theory templates,
and evaluate the biased mean values $\w_i^\mathrm{bias}$
of the nonperturbative physics parameters of interest.
Repeating this for several values of the bias parameters,
we can evaluate the required partial derivatives (``bias strengths'') $\partial \w_i^\mathrm{bias}/\partial B$
with respect to each of the bias parameters $B$ by finite differences,
leading to the results in \fig{bias_strength}.

\end{document}